\documentstyle[prd,aps,preprint,eqsecnum,psfig,floats]{revtex}
\def\bege{\begin{equation}}
\def\ende{\end{equation}}

\def\beq{\begin{equation}}
\def\eeq{\end{equation}}
\def\bea{\begin{eqnarray}}
\def\eea{\end{eqnarray}}

\begin{document}
\tighten
\draft

\title{Frame-Independence of Exclusive Amplitudes in the Light-Front 
Quantization}

\author{Chueng-Ryong Ji and Chad Mitchell}

\address{Department of Physics, North Carolina State University,
Raleigh, NC 27695-8202, USA}


\maketitle

\begin{abstract}

While the particle-number-conserving convolution formalism established in 
the Drell-Yan-West reference frame is frequently used to compute exclusive 
amplitudes in the light-front quantization,
this formalism is limited to only those frames where the
light-front helicities are not changed and the good (plus) component of the
current remains unmixed.
For an explicit demonstration of such criteria, we present the relations 
between the current matrix elements 
in the two typical reference frames used for calculations of the exclusive 
amplitudes, {\it i.e.} the Drell-Yan-West and Breit frames and investigate
both pseudoscalar and vector electromagnetic currents in detail. We find 
that the light-front helicities
are unchanged and the good component of the 
current does not mix with the other components of the current under the
transformation between these two 
frames. Thus, the pseudoscalar and vector form factors obtained by the 
diagonal convolution formalism in both frames must indeed be identical. 
However, such coincidence between the Drell-Yan-West and Breit frames
does not hold in general. We give an explicit example in which the 
light-front helicities are changed and the plus component of the current 
is mixed with other components under the change of reference frame.
In such a case, the relationship 
between the frames should be carefully analyzed before the established 
convolution formalism in the Drell-Yan-West frame
is used.
 
\end{abstract}
\pacs{ }

\section{Introduction}

The hadron phenomenology based on the equal-$\tau$ quantization takes 
great advantage of the 
Drell-Yan-West frame\cite{DYW}. 
In the $q^+ = 0$ Drell-Yan-West frame\cite{DYW},
one can derive a first-principle 
formulation for the exclusive amplitudes by judiciously 
choosing the good component of the light-front current. 
Due to the rational dispersion relation 
in the equal-$\tau$ formulation, one doesn't need to suffer from the 
complicated vacuum fluctuation.  The zero-mode contribution may also be 
avoided in the Drell-Yan-West frame by using the plus component of the current 
\cite{Ji}.  However, caution is needed in 
applying the established Drell-Yan-West formalism to other frames because 
in general the current components do mix under transformation of the 
reference-frame.  Furthermore, the Poincare algebra in the ordinary equal-$t$ quantization is
drastically changed in the light-front equal-$\tau$ quantization. 
The upshot of the difference may be summarized as the change of the
dynamical operators in the two quantizations. For example, the transverse 
rotation operator in the light-front quantization does not commute with 
the Hamiltonian possessing the dynamics.  
Likewise, the light-front helicities 
are not in general independent from the reference frame. 
Therefore, the adopted light-front formulation used in one particular
reference-frame to compute physical quantities is not gauranteed 
to work if another reference-frame is used. Consequently, it is important 
to carefully check if the same formulation can indeed be used in another 
frame for the correct calculation insuring the frame-independence of the 
physical quantities.

In this work, we present criteria which insure an identical formulation
in different frames of the light-front quantization. 
As an explicit example, we consider the well-known convolution formalism
for the exclusive processes established in the Drell-Yan-West($q^+=0$) frame
\cite{DYW} and investigate the criteria to insure the same formalism in 
different
reference-frames. In particular, we analyze the transformation of the
electromagnetic current elements and the light-front helicities between 
the Drell-Yan-West frame and the Breit($q^+=0$) frame\cite{Fred} which is 
another popular reference-frame used frequently by Frederico and 
collaborators for the calculation of exclusive amplitudes.

We found that four operations are needed to go from the 
Drell-Yan-West frame\cite{DYW} to the Breit frame\cite{Fred}. 
Combining the four operations, we find the equivalent single operation 
that can go back and forth between two frames: $U= \exp[i(\alpha
{\cal J}^{+} + \beta {\cal K}^{3})]$, where the coefficients $\alpha$ 
and $\beta$ of the 
transverse rotation ${\cal J}^+$ and the boost ${\cal K}^3$ are 
functions of $q^2$. It turns out that the 
single operation is dynamical (not kinematical) because transverse 
rotation is involved in the single transformation. This means that the 
two reference-frames are not only kinematically different but also 
dynamically inequivalent even though both frames have $q^+ = 0$. 
Thus, some new dynamics can come in by moving from the Drell-Yan-West frame 
to the Breit frame. This in principle raises a flag on various form 
factor calculations performed in the Breit frame because the simple 
convolution formulas that worked out quite well in the Drell-Yan-West frame may 
not apply in the Breit frame. We thus investigate in detail both 
pseudoscalar and vector form factors in these two typical light-front 
frames. We find that the helicities in the Drell-Yan-West frame are not 
changed in the Breit frame and thus the form factors obtained in the two 
frames must be indeed identical\footnote{ However, this doesn't mean that
any model calculations of form factors would give the same results in the
two frames because the phenomenological models may not be covariant.
For example, the angular condition for the vector meson form factors
is violated if the model calculation is not fully covariant.
Thus, our finding here provides an additional condition that the covariant
model must satisfy besides the angular condition.}. 
We find that the plus component of the Drell-Yan-West 
frame is proportional to the plus component of the Breit frame so that 
the same convolution formula can be used in both frames. This is a highly 
non-trivial result because such coincidence cannot be in general 
anticipated. We have also expanded the operation in a perturbative way.
Closed exact single operation results $U= \exp[i(\alpha {\cal J}^+ 
+\beta {\cal K}^3)]$ were compared with the perturbative results.

In general, however, the light-front helicities are changed and the
plus component of the current is mixed with other components under
the change of reference frame.
For example, the $V$-transformation was obtained as $V = \exp[i( 
-\alpha {\cal K}^- - \beta {\cal K}^3)]$ where 
${\cal K}^-$ is 
the light-front transverse boost so that $V$ commutes with the Hamiltonian.
This reveals that the obtained $V$-transformation is dynamically equivalent 
even though it is kinematically different. In this case, however, the 
plus component of the current in Drell-Yan-West frame is not proportional to 
that in the $V$-frame. Thus, the same convolution formula obtained in the 
Drell-Yan-West frame cannot be used in the $V$-frame.

In the following section (Section II), we present the four operations
needed to go from the Drell-Yan-West frame to the Breit frame and obtain
the explicit relations for the matrix elements of the current in the two 
frames, confirming the usual unitary tranformation rule of the vector 
operator. It is also assured that the pseudoscalar form factor must be 
identical in the two frames because the plus component is preserved.
In Section III, we apply the transformation to the helicities
and show that the vector form factors must be also identical in the two 
frames because the helcities are not changed in the two frames. 
In Section IV, the $V$-transformation
is presented to show that the usual convolution formula for the form factor
calculation in the Drell-Yan-West frame cannot be used even though the 
transformation commutes with the hamiltonian involving dynamics.
Conclusions and discussions follow in Section V. 
In the Appendix, the perturbative expansion
of the transformation between the frames using the 
Campbell-Baker-Hausdorff relation is compared with the exact closed form
of transformation.

\section{Transformation between Drell-Yan-West and Breit frames}

We begin by considering the absorption of a photon by a meson of mass
$M$.  The Drell-Yan-West reference frame is defined such that the initial 
four-momentum
of the particle, $p$, and the four-momentum of the photon, $q$, are given as 
follows:
\begin{eqnarray}
p &=& (p_{0},0,0,|{\bf p}|) \\ \nonumber
q &=& (\frac{-q^{2}}{2p^{+}},\sqrt{-q^{2}},0,\frac{q^2}{2p^{+}}).
\end{eqnarray}

We found the four transformations required to move from the Drell-Yan-West frame
to the Breit frame.  These are represented schematically in Fig. 1.
\begin{figure}
\centerline{
\psfig{figure=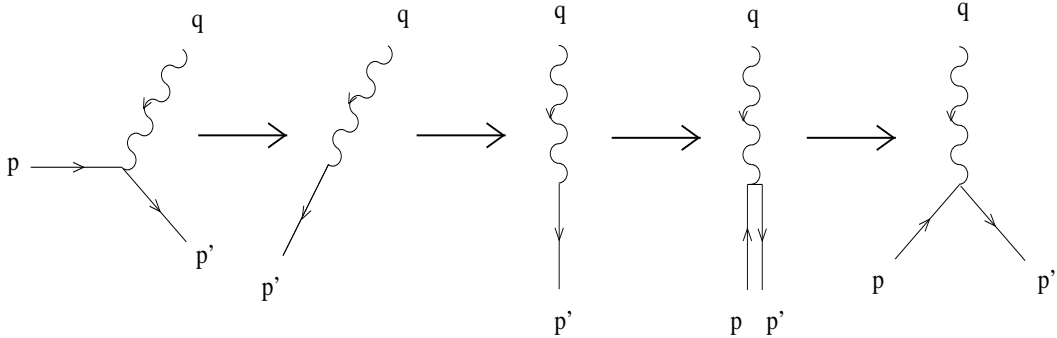,width=5.5in,height=2.5in}
}
\vspace{.1in}
\caption{The four transformations going from the Drell-Yan-West frame
to the Breit ($q^+=0$) frame.}
\label{DYW-B}
\end{figure}
The first transformation to the Breit frame is the boost in the z-direction 
which
eliminates the three-momentum of the meson.  The necessary boost
parameter $\omega$ must satisfy $\sinh{\omega}=\frac{|{\bf p}|}{M}$, generating the
Lorentz transformation
\begin{eqnarray}
\Lambda_{1} = \left(\begin{array}{cccc} \frac{E}{M} & 0 & 0 & 
-\frac{|{\bf p}|}{M} \\
0 & 1 & 0 & 0 \\ 0 & 0 & 1 & 0 \\ -\frac{|{\bf p}|}{M} & 0 & 0 & \frac{E}{M}
\end{array} \right).
\end{eqnarray}
The momenta in the resulting frame are $p = (M,0,0,0)$ and $q =
(\frac{-q^2}{2M},\sqrt{-q^2},0,\frac{q^2}{2M})$.

The second transformation is a rotation about the y-axis so that both the final
momentum of the massive particle and the momentum of the photon lie along the 
x-axis.  This requires that $\tan{\theta} = \frac{\sqrt{-q^2}}{2M}$.  
Throughout
this paper we will refer to the quantity $\frac{\sqrt{-q^2}}{2M}$ as 
$\kappa$. Therefore,
\begin{eqnarray}
\Lambda_{2} = \left(\begin{array}{cccc} 1 & 0 & 0 & 0 \\ 0 &
\frac{1}{\sqrt{1+{\kappa}^2}} & 0 & \frac{-\kappa}{\sqrt{1+{\kappa}^2}} \\ 0 
& 0 & 1 & 0
\\ 0 & \frac{\kappa}{\sqrt{1+{\kappa}^2}} & 0 & \frac{1}{\sqrt{1+{\kappa}^2}} 
\end{array} \right).
\end{eqnarray}
The initial momentum is unchanged, but the photon momentum is now 
$q = (2M {\kappa^2} , 2M {\kappa} {\sqrt{1+{\kappa^2}}},0,0)$ 
and the final 
momentum is
$p^\prime = (M + 2M {\kappa^2} ,2M {\kappa} {\sqrt{1+ {\kappa^2}}},0,0)$.

The next transformation is a boost by parameter $\omega_{2}$ in the 
x-direction
into a frame in which the incoming and outgoing momenta of the particle are 
equal and opposite,{\it i.e.}, ${\bf p} = -{\bf p}^\prime$.  Now 
$\sinh{\omega_{2}} = \kappa$, so that
\begin{eqnarray}
\Lambda_{3} = \left(\begin{array}{cccc} \sqrt{1+{\kappa}^{2}} & -\kappa & 
0 & 0 \\
-\kappa & \sqrt{1+{\kappa}^{2}} & 0 & 0 \\ 0 & 0 & 1 & 0 \\ 0 & 0 & 0 & 1
\end{array} \right).
\end{eqnarray}
Now the momenta are $p = (M\sqrt{1+{\kappa}^{2}},-M\kappa,0,0)$, $q =
(0,2M\kappa,0,0)$, and $p^{'} = (M\sqrt{1+{\kappa}^{2}},M\kappa,0,0)$.

The final transformation is a boost opposite the initial one by the parameter
$-\omega$ generating
\begin{eqnarray}
\Lambda_{4} = \left(\begin{array}{cccc} \frac{E}{M} & 0 & 0 & \frac{|{\bf 
p}|}{M} \\
0 & 1 & 0 & 0 \\ 0 & 0 & 1 & 0 \\ \frac{|{\bf p}|}{M} & 0 & 0 & \frac{E}{M}
\end{array} \right).
\end{eqnarray}
The resulting reference frame is known as the Breit frame, with new momenta:
$p = (p^{0}\sqrt{1+{\kappa}^{2}},-M\kappa,0,|{\bf p}| 
\sqrt{1+{\kappa}^{2}})$ and
$q = (0,2M\kappa,0,0)$.  If we define $Q = \sqrt{-q^{2}}$, the momenta in the
light-front formalism $k = (k^{+},k^{-},\bf k)$ become
\begin{eqnarray}
p &=&
(p^{+}_{i}\sqrt{1+{\kappa}^{2}},p^{-}_{i}\sqrt{1+{\kappa}^{2}},-\frac{Q}{2},0)
\\ \nonumber
q &=& (0,0,Q,0)
\\ \nonumber
p^{'} &=& 
(p^{+}_{i}\sqrt{1+{\kappa}^{2}},p^{-}_{i}\sqrt{1+{\kappa}^{2}},\frac{Q}{2},0)
\end{eqnarray}
where $p_{i}$ is the initial momentum of the massive particle in the
Drell-Yan-West frame.

The product of these four unitary transformations is a single unitary
transformation, $U$, which characterizes the relationship between the two 
frames.  To fully understand the invariance of electromagnetic form factors
under $U$, we must know how the matrix elements of the current transform.  We
explicitly verified that $U^{\dagger}I^{\mu}U = 
\Lambda^{\mu}_{\nu}I^{\nu}$, where
$\Lambda$ is the full Lorentz transformation matrix.  Applying this matrix to an
arbitrary current four-vector $I^\mu$ allows us to find the current-operator 
relations:
\begin{eqnarray}
U^{\dagger}I^{+}U &=& \sqrt{1+{\kappa}^{2}}I^{+}\\ \nonumber
U^{\dagger}I^{-}U &=& \frac{1}{\sqrt{1+{\kappa}^{2}}}I^{-} +
\frac{p^{-}_{i}\kappa}{M\sqrt{1+{\kappa}^{2}}}\left(\frac{p^{-}_{i}\kappa}{M}
I^{+} - 2I^{1}\right)\\ \nonumber
U^{\dagger}I^{1}U &=& I^{1} - \frac{p^{-}_{i}\kappa}{M}I^{+}\\ \nonumber
U^{\dagger}I^{2}U &=& I^{2}
\end{eqnarray}
Notice that the plus component of the current in Drell-Yan-West frame is 
proportional
to the plus component of the current in the Breit frame.  This 
correspondence is
necessary if the convolution formalism is to be valid in both frames.

We can use these relations to find a closed form for the generator of the
transformation.  First, note that the operator must have the form $U =
\exp{i\left({\alpha}{\cal J}^{+} + {\beta}{\cal K}^{3} + 
{\gamma}{\cal K}^{-}\right)}$, where we have
defined ${\cal J}^{+} = {\cal J}^{2} + {\cal K}^{1}$ and 
${\cal K}^{-} = {\cal K}^{1} - {\cal J}^{2}$ to be 
components of
angular momentum and boost on the light-front, respectively.  The coefficient 
$\gamma$ must be zero according to arguments presented in the Appendix.  
In Fig. 2, the first (dotted) and second (dashed-dotted) order perturbative 
expansions for $\gamma$ are
plotted versus the parameter $\kappa$ to demonstrate that the expansion
converges to $\gamma = 0$.
\begin{figure}
\centerline{
\psfig{figure=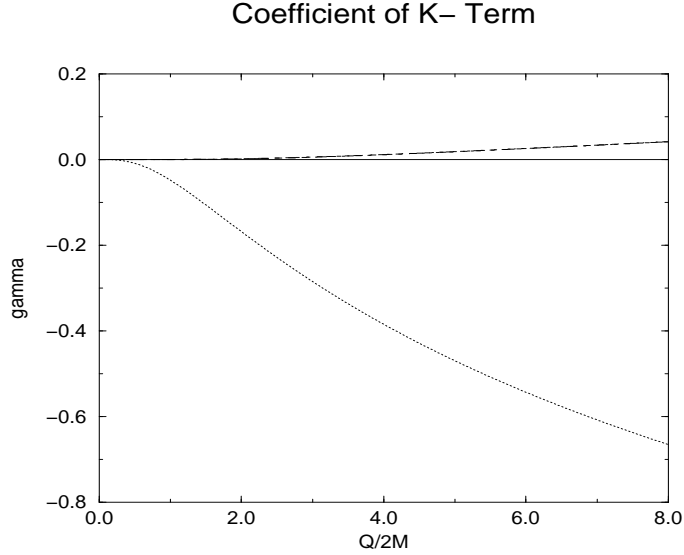,width=4.0in,height=3.0in}
}
\vspace{.1in}
\caption{Comparison between the perturbative results of $\gamma$ and the 
exact result $\gamma = 0$.}
\label{K-minus}
\end{figure}
Next, by using the Campbell-Baker-Hausdorff relation with the appropriate Poincare
algebra we find
\begin{eqnarray}
U^{\dagger}I^{+}U &=& I^{+}e^{\beta} \\ \nonumber
U^{\dagger}I^{1}U &=& I^{1} + 
I^{+}\frac{\alpha}{\beta}\left(e^{\beta}-1\right). \end{eqnarray}
Thus the full transformation from the Drell-Yan-West frame to the Breit 
frame can be represented by the unitary transformation
\begin{equation}
U = \exp{i\left({\alpha}{\cal J}^{+} + {\beta}{\cal K}^{3}\right)} 
\end{equation}
where
\begin{eqnarray}
\alpha &=&
\frac{\kappa}{1-\sqrt{1+{\kappa}^{2}}}\frac{p^{-}_{i}}{M}
\ln{\sqrt{1+{\kappa}^{2}}} \\ \nonumber
\beta &=& \ln{\sqrt{1+{\kappa}^{2}}}.
\end{eqnarray}
This result was checked explicitly by both construction of the 
corresponding Lorentz
transformation matrix and comparison with the perturbative expansion of
$e^{i\omega{{\cal K}^{3}}}e^{-i\omega_{2}{{\cal K}^{1}}} 
e^{-i\theta{{\cal J}^{2}}}e^{-i\omega{{\cal K}^{3}}}$ as 
presented in the Appendix.
Plots of the coefficients $\alpha$ and $\beta$ are included for comparison with
the perturbative expansion with $\omega = 0$ in Figs. 3 and 4.  Solid 
lines correspond to exact
solutions; dotted lines are first-order expansions; and dashed-dotted 
lines are up to third-order expansions. Note that the second-order terms are 
absent as shown in the Appendix if $\omega = 0$ (See Eq.(A.4)).
The dashed-dotted line is indistinguishable from the solid line in Fig. 4.

\begin{figure}
\centerline{
\psfig{figure=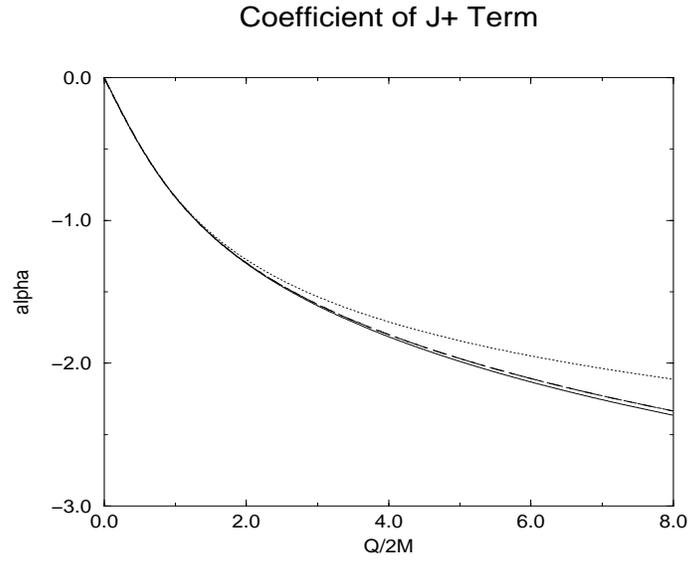,width=4.0in,height=3.0in}
}
\vspace{.1in}
\caption{Comparison of $\alpha$ between the perturbative results and
the exact result.}
\label{J-plus}
\end{figure}
\begin{figure}
\centerline{
\psfig{figure=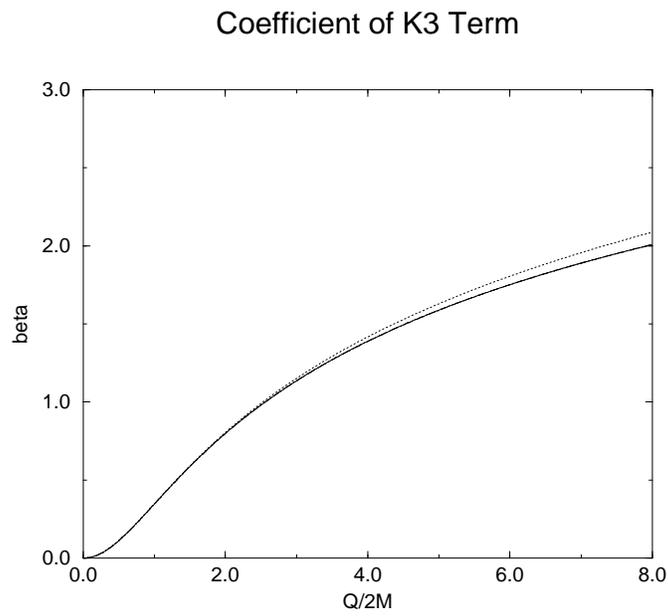,width=4.0in}
}
\vspace{.1in}
\caption{Comparison of $\beta$ between the perturbative results and
the exact result.}
\label{K-3}
\end{figure}

To verify the invariance of the pseudoscalar form factor under this
transformation, consider the relation $<p'|I^{\mu}|p> = (p+p')^{\mu} F$, 
where the
matrix element and the momenta are defined in the Drell-Yan-West frame.   In 
the Breit frame this becomes $<p'_D|U^{\dagger}I^{\mu}U|p_D> =
\Lambda^{\mu}_{\nu}<p'_D|I^{\nu}|p_D> = \Lambda^{\mu}_{\nu}(p+p')^{\nu}_D F =
(p+p')^{\mu}_{B}F$.  Since the relation must hold in any reference frame, 
we see that the form factor
$F$ is invariant under Lorentz transformations.  This can be easily 
confirmed for
the $U$ transformation by using the current-operator relations 
given by Eq.(2.7).

More importantly, the Drell-Yan-West particle-number-conserving convolution
formalism will be valid in the Breit frame.  This is true since
the plus component of any four-vector ({\it i.e.}, current) is unmixed 
with other components under this change of reference frame indicating  
$q^{+}=0 \rightarrow q^{+}=0$. Thus, the non-valence (pair-creation) 
diagrams remain excluded.

\section{Transformation of Helicities and the Equivalence of Vector Form 
Factors}

The invariance of vector or spin-one form factors is nontrivial due to the
frame-dependence of light-front helicities.  In this section we find that
light-front helicity eigenvalues are unchanged under transformation from the
Drell-Yan-West to the Breit reference frame.

The polarization vectors for light-front helicity eigenstates in the
Drell-Yan-West frame are given by
\begin{eqnarray}
\epsilon(0) &=& \frac{1}{M}(p^{+}_{i},-p^{-}_{i},0,0)\\ \nonumber
\epsilon(+1) &=& -\frac{1}{\sqrt{2}}(0,0,1,i)\\ \nonumber
\epsilon(-1) &=& \frac{1}{\sqrt{2}}(0,0,1,-i)\\ \nonumber
\epsilon^{'}(0) &=& 
\frac{1}{M}(p^{+}_{i},-p^{-}_{i}+\frac{Q^{2}}{p^{+}_{i}},Q,0)\\ \nonumber
\epsilon^{'}(+1) &=& -\frac{1}{\sqrt{2}}(0,\frac{2Q}{p^{+}_{i}},1,i)\\ 
\nonumber
\epsilon^{'}(-1) &=& \frac{1}{\sqrt{2}}(0,\frac{2Q}{p^{+}_{i}},1,-i)
\end{eqnarray}
while the corresponding vectors in the Breit frame are:
\begin{eqnarray}
\epsilon(0) &=&
\frac{1}{M}(p^{+}_{i}\sqrt{1+{\kappa}^{2}},
p^{-}_{i}\frac{{\kappa}^{2}-1}{\sqrt{1+{\kappa}^{2}}},-\frac{Q}{2},0)
\\ \nonumber
\epsilon(+1) &=&
-\frac{1}{\sqrt{2}}(0,-\frac{Q}{p^{+}_{i}\sqrt{1+{\kappa}^{2}}},1,i)
\\ \nonumber
\epsilon(-1) &=& 
\frac{1}{\sqrt{2}}(0,-\frac{Q}{p^{+}_{i}\sqrt{1+{\kappa}^{2}}},1,-i)
\\ \nonumber
\epsilon^{'}(0) &=& 
\frac{1}{M}(p^{+}_{i}\sqrt{1+{\kappa}^{2}},
p^{-}_{i}\frac{{\kappa}^{2}-1}{\sqrt{1+{\kappa}^{2}}},\frac{Q}{2},0)
\\ \nonumber
\epsilon^{'}(+1) &=& 
-\frac{1}{\sqrt{2}}(0,\frac{Q}{p^{+}_{i}\sqrt{1+{\kappa}^{2}}},1,i)
\\ \nonumber
\epsilon^{'}(-1) &=& 
\frac{1}{\sqrt{2}}(0,\frac{Q}{p^{+}_{i}\sqrt{1+{\kappa}^{2}}},1,-i).
\end{eqnarray}
In each frame, the polarization vector representing a particular helicity
eigenstate is determined by applying the three constraints
\begin{eqnarray}
\epsilon(p,\lambda)\cdot{p} &=& 0
\\ \nonumber
\epsilon^{*}(p,\lambda)\cdot\epsilon(p,\lambda) &=& -\delta_{\lambda\lambda'}
\\ \nonumber
\sum_{\lambda}\epsilon^{\mu}(p,\lambda)\epsilon^{\nu}(p,\lambda) &=& -g^{\mu\nu}
+ \frac{p^{\nu}p^{\mu}}{M^{2}}
\end{eqnarray}
and choosing $\epsilon(p,0)$ so that its plus and perpendicular components are
proportional to the plus and perpendicular components of the momentum $p$.  In 
general, any polarization vector in one frame transforms to a superposition of 
polarization vectors in a second reference frame.  Thus, light-front helicities
are not Lorentz invariant.  Applying the transformations described in the
previous section to the six polarization vectors in the Drell-Yan-West frame,
however, we obtain exactly the corresponding six vectors in the Breit frame.  
This implies that a state of helicity $\lambda$ in the Drell-Yan-West frame 
transforms to a state of the same helicity $\lambda$ in the Breit frame.  

One consequence of this special relationship between the Drell-Yan-West and
Breit reference frames is that the form of the angular condition, which is
used to check the accuracy of model vector meson wavefunctions, is identical in
both reference frames. The angular condition in the Drell-Yan-West frame is 
given by \cite{GK}

\begin{equation}
\Delta(Q^2) =
(1+2{\kappa}^2)F^{+}_{++}+F^{+}_{+-}-\kappa{\sqrt{8}}F^{+}_{+0}-F^{+}_{00} 
= 0,
\end{equation}
where $F^{+}_{\lambda\lambda'} = <p',\lambda'|I^{+}|p,\lambda>$.
To see how Eq.(3.4) holds also in the Breit frame, first note that 
$<p'_B,\lambda'_B|I^{\mu}|p_B,\lambda_B> =
<p'_{D},\lambda'_D|U^{\dagger}I^{\mu}U|p_{D},\lambda_D>$.  Thus, each matrix 
element transforms according to the current-operator relations 
(see Eq.(2.7)) presented in the last
section. Since the helicity is invariant under this particular 
transformation, the matrix elements in the Breit frame are given by
\begin{eqnarray}
<p'_{B},\lambda'_{B}|I^{\mu}|p_{B},\lambda_{B}> &=&
\sum_{m,n}<p',\lambda'(n)|U^{\dagger}I^{\mu}U|p,\lambda(m)>c_{m}c_{n} 
\\ \nonumber
&=& 
<p'_D,\lambda'_D|U^{\dagger}I^{\mu}U|p_D,\lambda_D> =
\Lambda^{\mu}_{\nu}<p'_D,\lambda'_D|I^{\nu}|p_D,\lambda_D>.
\end{eqnarray}
This shows that $F^{+}_{\lambda\lambda'} \rightarrow 
\sqrt{1+{\kappa}^{2}}F^{+}_{\lambda\lambda'}$ under the transformation 
from the Drell-Yan-West frame to the Breit frame. Therefore, Eq.(3.4)  must 
also hold in the Breit frame.
Similarly, one can show not only that $q^+ = 0 \rightarrow q^+ = 0$
but also that the relations between the vector form factors 
$\{F_1,F_2,F_3\}$ 
and $F^{+}_{\lambda\lambda'}$ remain the same in both frames.
This is a remarkable feature of the two frames which results in 
the invariance of the convolution formalism.

Furthermore, the matrix elements are related to the vector form factors as 
follows \begin{eqnarray}
<p',\lambda'|I^{\mu}|p,\lambda> =
\epsilon^{*'}_{\alpha}\epsilon_{\beta}[-g^{\alpha\beta}(p+p')^{\mu}F_{1} +
(g^{\mu\beta}q^{\alpha}-g^{\mu\alpha}q^{\beta})F_{2} +
\frac{q^{\alpha}q^{\beta}(p+p')^{\mu}F_{3}}{2M^{2}}].
\end{eqnarray}
Under Lorentz transformations all scalar products on the right remain
invariant, so that each term on the right-hand side transforms as the 
$\mu$ component of a momentum or polarization four-vector times a form 
factor. This means that in the Briet frame the matrix elements can be written
as
\begin{eqnarray}
<p'_B,\lambda'_B|I^{\mu}|p_B,\lambda_B> = \sum_i {\rho^{\mu}_B}_i F^B_i,
\end{eqnarray}
where  ${\rho^{\mu}_B}_i (i=1,2,3)$ is the coefficient of $F_i$
(see Eq.(3.6)) in the Breit frame.
According to Eq.(3.5), the left-hand side of Eq.(3.7) is also
equal to 
\begin{eqnarray}
\Lambda^{\mu}_{\nu}<p'_D,\lambda'_D|I^{\nu}|p_D,\lambda_D>
&=& \Lambda^{\mu}_{\nu} \sum_i {\rho^{\mu}_D}_i F^D_i \\
\nonumber
&=& \sum_i {\rho^{\mu}_B}_i F^D_i.
\end{eqnarray}
Therefore, $\sum_i {\rho^{\mu}_B}_i F^B_i =\sum_i {\rho^{\mu}_B}_i F^D_i$.
This can only be true if all three form factors $F_{1},F_{2},F_{3}$ are 
identical in both frames. 

\section{The $V$-Transformation}

It is well-known that helicities in the traditional equal-time quantization are
frame dependent.  We expect a similar phenomenon on the light-front.  In order
to demonstrate that light-front helicities are not frame-invariant in general,
consider the following example.  Define a transformation $V$ which is 
identical to the 
previous transformation, $U$, except that each boost is reversed; i.e.,
$\omega \rightarrow -\omega$ and $\omega2 \rightarrow -\omega2$.  
Under this transformation we find the current-operator relations
\begin{eqnarray}
V^{\dagger}I^{-}V &=& \sqrt{1+{\kappa}^{2}}I^{-}\\ \nonumber
V^{\dagger}I^{+}V &=& \frac{1}{\sqrt{1+{\kappa}^{2}}}I^{+} +
\frac{p^{-}_{i}\kappa}{M\sqrt{1+{\kappa}^{2}}}\left(\frac{p^{-}_{i}\kappa}{M}
I^{-} + 2I^{1}\right)\\ \nonumber
V^{\dagger}I^{1}V &=& I^{1} + \frac{p^{-}_{i}\kappa}{M}I^{-}\\ \nonumber
V^{\dagger}I^{2}V &=& I^{2}.
\end{eqnarray}
Note that the plus component of the current is not proportional to the plus
component of the current in the Drell-Yan-West frame.
Now applying the same 
constraints (3.3)
as before, we obtain the polarization vectors for the initial meson:
\begin{eqnarray}
\epsilon(0) &=& 
\frac{1}{M}(k^{+},\frac{{k_\perp}^{2}-M^{2}}{k^{+}},{k_\perp},0)
\\ \nonumber
\epsilon(+1) &=& -\frac{1}{\sqrt{2}}(0,\frac{2{k_\perp}}{k^{+}},1,i)
\\ \nonumber
\epsilon(-1) &=& \frac{1}{\sqrt{2}}(0,\frac{2{k_\perp}}{k^{+}},1,-i),
\end{eqnarray}
where the parameters $k^{+},{k_\perp}$ are components of the fermion's
light-front momentum in the new $V$-frame which is given by
\begin{eqnarray}
k = (\frac{p^{-}}{M^{2}{\sqrt{1+{\kappa}^{2}}}} \left[(p^{+})^{2} + 
{\kappa}^{2} (p^{-})^{2}\right], 
p^{-}{\sqrt{1+{\kappa}^{2}}}, \frac{\kappa (p^{-})^{2}}{M}, 0). 
\end{eqnarray}
Similarly, the components of the photon's light-front momentum $k_{\gamma}$ in 
the $V$-frame are
\begin{equation}
k_{\gamma} = (\frac{1}{{p^{+}}M^{2}{\sqrt{1+{\kappa}^{2}}}} \left[M^{2} + 
{\kappa}^{2}
(p^{-})^{2}\right],\frac{Q^2}{p^{+}}\sqrt{1+{\kappa}^2},\frac{Q}{p^{+}}(2{x^2}p^{-}
+ p^{+}),0).
\end{equation}
The transformation $V$ can be represented as the unitary operator $V =
\exp{i[-{\alpha}{\cal K}^{-}-{\beta}{\cal K}^{3}]}$, where 
$\alpha$ and $\beta$ were presented in section II.

Transforming the Drell-Yan-West polarization vectors according to $V$ yields a 
set of
polarization vectors in this new $V$-frame.  None of these polarization 
vectors,
however, represents a helicity eigenstate in this frame.  We obtain, for
example, that
\begin{eqnarray}
\epsilon(0)^{+} = \frac{p^{-}}{M^{3}\sqrt{1+{\kappa}^{2}}}\left[(p^{+})^{2} - 
{\kappa}^{2}(p^{-})^{2}\right].
\end{eqnarray}
The correct plus component of the zero-helicity polarization vector is
$\frac{k^{+}}{M}$ from Eq.(4.2) or 
\begin{eqnarray}
\epsilon(0)^{+} = \frac{p^{-}}{M^{3}\sqrt{1+{\kappa}^{2}}}\left[(p^{+})^{2} + 
{\kappa}^{2}(p^{-})^{2}\right],
\end{eqnarray}
according to Eq.(4.3).
The subtle sign difference is a consequence of changing the sign of each boost 
in the transformation.  For a stronger example, consider the transverse
polarization vectors.  We obtain by transforming the Drell-Yan-West vectors
\begin{eqnarray}
\epsilon(+1)^{+} &=& 
-\frac{1}{\sqrt{2}}(\frac{2{\kappa}p^{-}}{M\sqrt{1+{\kappa}^{2}}})
\\ \nonumber
\epsilon(-1)^{+} &=& 
\frac{1}{\sqrt{2}}(\frac{2{\kappa}p^{-}}{M\sqrt{1+{\kappa}^{2}}}).
\end{eqnarray}
These terms must be zero, as presented in (4.2), to satisfy the necessary
constraints in the $V$-frame.  Thus, light-front helicity is not invariant 
under this $V$-transformation.  This implies that the convolution formulation 
used in the
Drell-Yan-West frame cannot be used in the $V$-frame.  As a result, we 
expect a different form for the angular condition in this frame.

\section{ Conclusion }

We applied the four operations in the calculation of the pseudoscalar form 
factor and found that the form factor can be identically obtained in the
Drell-Yan-West and Breit reference frames using the simple convolution 
formalism as long as the plus 
component of the current is used.  We also applied the four operations in the 
calculation of spin-1 (vector) form factors, $F_1, F_2, F_3 (G_E, G_M, 
G_Q)$. We found that 
the light-front helicities become identical in the two frames even though the 
light-front helicities are in general frame dependent.
Thus, all three form factors obtained by the particle-number-conserving 
convolution formalism must be identical in 
the two frames confirming the correctness of previous applications to 
the vector meson form factors in the Breit frame.
We also find that the angular condition is identical
in the two frames. This is a 
remarkable result because it works only in very limited special frames. 
Thus, the Drell-Yan-West and Breit frames can be regarded as such 
special frames. 
However, such coincidence does not generally hold for other reference frames.
One needs thus to investigate the frame-independence of form factors with 
care before relying on a particular reference frame. 
In summary, the two typical frames (Drell-Yan-West and Breit) are 
special in the light-front computation since the plus components of the 
current in the two frames are proportional to each other and the 
light-front helicities are equivalent in these two frames. Thus, the angular 
condition is also identical and the same convolution formula obtained in the 
Drell-Yan-West frame can equivalently be used in the Breit frame.  
However, in other frames, caution is needed in using the Drell-Yan-West 
convolution formalism. 

\acknowledgements
\noindent
This work was supported in part by a grant from the US Department of
Energy.

\newpage
\noindent
\appendix
\setcounter{section}{0}
\setcounter{equation}{0}
\setcounter{figure}{0}
\renewcommand{\theequation}{\mbox{A\arabic{equation}}}
\section{ Comparison with Perturbative Expansion Using CBH 
Relations}

According to the Campbell-Baker-Hausdorff Theorem, if $e^{A}e^{B} = 
e^{C}$
then $C$ can be expressed as an expansion in terms of commutators.  The 
first few terms are
\begin{eqnarray}
C = A + B + \frac{1}{2}[A,B] + \frac{1}{12}([A,[A,B]]-[[A,B],A]) + \cdots.
\end{eqnarray}
Applying this theorem three times allows us to combine the unitary
transformation
$e^{i\omega{{\cal K}^3}} 
e^{-i\omega_{2}{{\cal K}^1}}e^{-i\theta{{\cal J}^2}}
e^{-i\omega{{\cal K}^3}}$
into a single exponential, providing us with a perturbative expansion for the
generator of the $U$-transformation.
\begin{eqnarray}
C &=&
\frac{1}{2}{\cal J}^{+} 
(a+b)(1-\frac{1}{2}\omega+\frac{1}{12}{\omega}^{2}+
\frac{1}{12}\omega{c}) +
\frac{1}{2}{\cal K}^{-}(a-b)(1-\frac{1}{2}\omega+\frac{1}{12}{\omega}^{2}+
\frac{1}{12}\omega{c}) 
\\ \nonumber
&+&
{\cal K}^{3}(c-\omega-\frac{1}{12}\omega(a^{2}-b^{2})) + \cdots.
\end{eqnarray}
Where $a, b,$ and $c$ are given up to fourth order as
\begin{eqnarray}
a &=& -\omega_{2} + \frac{1}{2}\omega\theta
+\frac{1}{12}\omega_{2}(-{\omega}^{2}+{\theta}^{2}) +
\frac{1}{12}({\omega_{2}}^{2}\omega\theta)
\\ \nonumber
b &=& -\theta + \frac{1}{2}\omega\omega_{2}
-\frac{1}{12}\theta({\omega}^{2}+{\omega_{2}}^{2})
\\ \nonumber
c &=& \omega + \frac{1}{2}\theta\omega_{2}
+\frac{1}{12}\omega(-{\theta}^{2}+{\omega_{2}}^{2}).
\end{eqnarray}
Suppose that $\omega = 0$.  Then, keeping up to third order, we have
\begin{eqnarray}
C =
\frac{1}{2}{\cal J}^{+} 
(-\omega_{2}-\theta+\frac{1}{12}{\theta}^{2}\omega_{2}-
\frac{1}{12}{\omega_{2}}^{2}\theta)+
\frac{1}{2}{\cal K}^{-}
(-\omega_{2}+\theta+\frac{1}{12}{\theta}^{2}\omega_{2}+
\frac{1}{12}{\omega_{2}}^{2}\theta)+
\frac{1}{2}{\cal K}^{3}\omega_{2}\theta.
\end{eqnarray}
Using the facts that $\tan{\theta} = \kappa$ and $\sinh{\omega_{2}} = 
\kappa$, we can
examine the coefficient of each operator versus the parameter $\kappa$. 
In Figs.2-4, each coefficient corresponding to $\alpha, \beta$, and
$\gamma$ in Eq.(A.4) are compared with the closed form presented earlier
(See Eq.(2.10)).  
For small values of $\kappa$ the closed form and the expansion agree well, 
and their values
diverge slowly with increasing $\kappa$ as expected.  The coefficient of the
${\cal K}^{-}$ operator approaches zero for a given $\kappa$ as the 
order of
approximation increases.
One can indeed verify $\gamma = 0$ because we have already shown that
the plus component of current in the Drell-Yan-West frame 
must be
proportional to the plus component of current in the Breit frame (See 
Eq.(2.7)).  
Since the commutator $[I^{+},{\cal K}^{-}] = 2iI^{1}$, including a 
${\cal K}^{-}$ term in the
transformation would contradict Eq.(2.7).
Thus, the coefficient $\gamma$ must be zero.

\end{document}